# Rapid Fluctuations in the Lower Solar Atmosphere

J. K. Lawrence, A.C. Cadavid, D. J. Christian

Department of Physics and Astronomy, California State University, Northridge

18111 Nordhoff Street, Northridge, CA 91330-8268, USA

john.lawrence@csun.edu

D. B. Jess, M. Mathioudakis

Astrophysics Research Centre, School of Mathematics and Physics,

Queen's University Belfast, Belfast, BT7 1NN, Northern Ireland, U. K.

Proposed Running Title: **Rapid Fluctuations in the Solar Atmosphere**

# ABSTRACT

The Rapid Oscillations in the Solar Atmosphere (ROSA) instrument reveals solar atmospheric fluctuations at high frequencies. Spectra of variations of the G-band intensity ($I_G$) and CaII K-line intensity ($I_K$) show correlated fluctuations above white noise to frequencies beyond 300 mHz and 50 mHz, respectively. The noise-corrected G-band spectrum for $f$ = 28 - 326 mHz shows a power law with exponent -1.21 ± 0.02, consistent with the presence of turbulent motions. G-band spectral power in the 25 - 100 mHz ("UHF") range is concentrated at the locations of magnetic bright points in the intergranular lanes and is highly intermittent in time. The intermittence of the UHF G-band fluctuations, shown by a positive kurtosis $\kappa$, also suggests turbulence. Combining values of $I_G$, $I_K$, UHF power, and $\kappa$, reveals two distinct states of the solar atmosphere. State 1, including almost all the data, is characterized by low $I_G$, $I_K$, and UHF power and $\kappa \approx 6$. State 2, including only a very small fraction of the data, is characterized by high $I_G$, $I_K$, and UHF power and $\kappa \approx 3$. Superposed epoch analysis shows that the UHF power peaks simultaneously with spatio-temporal $I_G$ maxima in either state. For State 1, $I_K$ shows 3.5 min chromospheric oscillations with maxima occurring 21 s after $I_G$ maxima implying a 150 - 210 km effective height difference. However, for State 2 the $I_K$ and $I_G$ maxima are simultaneous; in this highly magnetized environment sites of G-band and K-line emission may be spatially close together.

Key words Sun: photosphere – Sun: chromosphere – Sun: oscillations

# 1. Introduction

As solar observations with increasing spatial and temporal resolution become available, the search continues for an energy supply to balance radiative losses in the chromosphere. Of renewed interest are contributions from high frequency magnetoacoustic fluctuations. Bello Gonzalez et al. (2010) using Doppler images from the IMaX two-dimensional spectropolarimeter aboard SUNRISE, with spatial resolution 100-130 km, found an energy flux of ~7000 W/m$^2$ in the 5.2 - 10 mHz range. This is half of the highest radiative loss estimate of 14,000 W/m$^2$ (Anderson & Athay 1989). High-frequency ($f$ >10 mHz) magnetoacoustic waves also may play an important role in chromospheric heating. Hasan, et. al., (2005) and Hasan & van Ballegooijen (2008) analyzed photospheric motions and found it reasonable to expect turbulent velocity fluctuations in intergranular lanes with timescales ~20 sec and that these fluctuations might buffet the footpoints of magnetic flux tubes producing upward propagating hydromagnetic waves. They therefore proposed that chromospheric heating could occur due to waves with frequencies greater than 10 mHz.

Frequencies as high as 100 mHz have been detected in TRACE data (DeForest 2004), but the detection of high frequency waves has been difficult because of atmospheric seeing and space-based instrumental and telemetry limitations (e.g. Fossum and Carlsson 2006). Now, however, the required observations with higher spatial and time resolution are more readily available (Carlsson, et al. 2007) with the advent of faster CCDs and improved image reconstruction.

Here we use data obtained with the Rapid Oscillations in the Solar Atmosphere (ROSA) instrument (Jess et al. 2010) at the Dunn Solar Telescope (DST) to identify high frequency fluctuations both in the photosphere and upper photosphere/chromosphere. We investigate the temporal and spatial properties of the fluctuations and test them against the properties predicted in the model of Hasan and van Ballegooijen (2008). We do not anticipate wave-like oscillations at high frequencies, but rather intermittent and turbulent fluctuations in the photosphere, manifested by power-law spectra.

## 2. Observations

Images were obtained with the ROSA instrument 2009 May 28 (Jess, et al. 2010). A sequence of exceptionally good seeing and alignment covered 32 Mm square at disk center and spanned 32.34 minutes. We use observations taken in the G-band of the CH radical at 4305.5 Å, bandpass 9.2 Å, with exposure times 15 msec at a cadence of 30.3 frames per second. These were reconstructed to a cadence of 0.525 sec. We also use images in the Ca II K-line core at 3933.7Å, bandpass 1.0 Å, with cadence 4.2 sec. High-order adaptive optics (AO) corrected wave front deformations in real time (Rimmele 2004), and the images underwent photometrically accurate speckle reconstruction (Weigelt & Wirnitzer 1983; Wöger et al. 2008) to reach spatial resolution 100 km (Figure 1). In the present work the G-band images were averaged to a 1.05 sec cadence for calculating the spectrum and a 4.2 sec cadence otherwise.

## 3. Analysis and Results

### 3.1 Identification of Ultra High Frequencies

We calculated the complex Morlet wavelet transform $W(t, \tau)$ of the intensity time signals at every spatial location and each time step. The spectral power is $P = |W(t, \tau)|^2$, and $\tau$ is the period (Torrence & Compo 1998). We averaged over space-time voxels excluding those within 3.5 min from the ends of the data string. This gives the raw spectra in Figure 2 shown as a thin solid line for G-band and a thin dashed line for K-line. The K-line and G-band fluctuations at cadences of 4.2 sec and 1.05 sec, have Nyquist frequencies of 120 mHz and 480 mHz, respectively. Uncorrelated (white) noise produces a flat spectrum. The raw G-band spectrum in Figure 2 flattens above 300 mHz, and the K-line above 50 mHz. Since the white noise must be uncorrelated with any signal, the noise contribution to the spectrum can be subtracted off (Press, et al. 1988). We model the G-band spectral power for high frequencies as a power law in frequency with exponent -$\alpha$ plus a constant chosen to minimize the uncertainty of $\alpha$. Then a fit over the range $28 < f < 326$ mHz (more than a decade in frequency) gives $\alpha = 1.21 \pm 0.02$. The corrected G-band spectrum is shown in Figure 2 as a thick solid line. The power law fit is shown by a dotted line. The similarly corrected K-line spectrum has a maximum corresponding to 3.5-minute oscillations, then falls off with a power law spectrum with index $\alpha \approx 4$. This is shown as a heavy dashed line.

To study their physical origin, we integrated the G-band spectra at each space-time voxel over the 25 - 100 mHz band, which we arbitrarily call the "UHF" band, and averaged over time. The UHF power is widely distributed in area, but strong concentrations (Figure 1) appear at the locations of G-band bright points in the intergranular lanes. We associate the bright points with magnetic flux tubes and refer to them as "magnetic bright points" (MBPs). The UHF concentrations also align closely with the bright cores of the K-line emission (Figure 1).

Wavelets can analyze signals in both frequency and time. Figure 3 shows a time plot of the UHF G-band wavelet power at the spatial location of an MBP. The highly intermittent variations are clear. If we view in a movie sequence the individual frames that are averaged in Figure 1, the UHF concentrations come and go intermittently. See the animation in the online version of the article.

## 3.2 UHF and Intensity

We studied various quantities as functions of G-band Intensity ( $I_G$). $I_G$ was divided into bins of width 0.01 in arbitrary units. Then, sampling all space-time voxels in the corresponding data cube gives the mean of the K-line intensity ($I_K$) in each bin. After multiplication by an arbitrary factor for convenience of display, the solid curve in Figure 4 shows the mean $I_K$ as a function of $I_G$. There is a lower plateau for $0.6 \leq I_G \leq 1.4$, a transition, and then a higher plateau for $I_G \geq 1.65$. For convenience, we call these cases "State 1" and "State 2," respectively. Although the transition of the mean $I_K$ appears continuous, a plot of the full histograms indicates that in the transition the population of points with lower $I_K$ decreases as $I_G$ increases above 1.4 while a population of points with greater $I_K$ appears and increases in number. This suggests two distinct states of the solar atmosphere. State 1 includes 98.6% of all data voxels, State 2 State 2 only 0.15%.

We next investigated the dependence on $I_G$ of the UHF G-band power. The wavelet spectra are normalized by the square of the mean $I_G$ at the spatial location where the spectrum is calculated. This is to offset bias in the UHF spectra caused by the magnitude of the intensity. The dashed line in Figure 4 shows a plateau in State I, then a steep linear transition, followed by a reduced-slope linear dependence in State 2. This again points to two distinct states. In contrast, the 3 - 5 mHz power spectrum (not shown) does not have a well-defined transition, although it does increase for larger $I_G$.

Still more can be learned about the processes underlying the fluctuations by studying the shapes of their distribution functions. The kurtosis κ of a distribution is its fourth moment about the mean divided by the second moment squared, all minus 3. With this definition the κ of a Gaussian distribution is zero. Distributions with κ < 0 have truncated tails, such as for harmonic oscillations. Distributions with $\kappa > 0$ have heavy tails. These indicate intermittency and are found, for example, in turbulent motions. In Figure 4 we have plotted the kurtosis of the $I_G$ fluctuations, after filtering to the UHF bandpass, and averaging in each $I_G$ bin. For State 1, $\kappa$ lies between values of 5 and 6. Then there is a clear transition, and in State 2 $\kappa$ drops to a value around 3. We thus find that the UHF fluctuations are intermittent in both States 1 and 2. In State 1 the fluctuations are highly intermittent and suggest the presence of turbulent motions. In State 2 the fluctuations are still intermittent, although less so. These may reflect the buffeting of magnetic flux elements by turbulence. In the case of G-band 3 - 5 mHz fluctuations (not shown) we find that $\kappa \approx -1$ over the whole range of $I_G$. This indicates a truncated distribution and implies harmonic oscillations.

### 3.3 Superposed Epoch Analysis

The relative timing of fluctuations in $I_G$ and $I_K$ were best studied by superposed epoch analysis (Lühr et al., 1998). This simple method is used when there are many observations of a certain kind of event and one wants to investigate the real responses minus any noise. By averaging over many cases the real signal remains while the noisy contributions are suppressed. In the present example we first locate local spatio-temporal maxima of $I_G$. In each such case we record the $I_G$ time series for 4 minutes before and after the maximum, which is taken as the zero point. We record the simultaneous values of the $I_K$ time series, and the G-band UHF power series, each with the same zero point. We repeat this for all $I_G$ maxima meeting some criterion, such as that the maximum lies in State 1, and then average the $I_G$, $I_K$ and the UHF spectral power series within each time bin 4.2 sec apart. Overall, we find 446,431 qualifying maxima in State 1 and 5,585 in State 2.

Figure 5 shows the resulting plots made separately for State 1 and for State 2. The two plots for UHF power each have a peak at t = 0, coincident with the local $I_G$ maxima. The UHF spectra have been normalized with respect to the mean $I_G$ at each spatial location. In State 1 the $I_K$ plot shows oscillations with period about 3.5 min, consistent with the chromospheric oscillations. The maximum of $I_K$ is delayed after the $I_G$ maximum by 21 sec. For State 2, no $I_K$ oscillations are apparent in our 8 minute window, and the $I_K$ maximum is simultaneous with the $I_G$ maximum.

## 4. Summary and Discussion

We identified significant power for fluctuations in $I_K$ beyond 50 mHz, and in $I_G$ beyond 300 mHz. Jess et al. (2007) identified frequencies up to 50 mHz in G-band and Hα observations made with the Rapid Dual Imager, a predecesor of ROSA.

After correction for white noise, we obtained power-law spectra. For $28 < f < 326$ mHz the G-band spectrum has exponent α = 1.21 ± 0.02. This indicates scale invariance and, because α is different from 0 (white noise) and 2 (Brownian motion) it points to a nonlinear process, such as chaos or turbulence (Bak, et al. 1987). Using $I_G$ as a proxy for the strength of the magnetic field (Schüssler et al. 2003), we found that the UHF G-band fluctuations were spatially associated with MBPs in the intergranular lanes as well as with K-line bright cores. Lawrence, et al. (2001) had earlier found that the spatial and temporal scaling of MBP motions resembled the walks of imperfectly correlated tracers of turbulent fluid flow in intergranular lanes.

With corrected ground-based observations come concerns about residual seeing effects. In this case the atmospheric seeing was optimal, and the application of high-order AO correction, photometrically accurate speckle reconstruction, and a dense 40x40 de-stretching grid, should remove all residual feature displacements. The KISIP speckle reconstruction code (Wöger, et al. 2008) has been checked for photometric accuracy against contemporaneous space-based Hinode SOT/FG images.

We have checked the possibility that the association of UHF power with the small MBPs might be a feature-size-dependent scintillation effect. The increase of high-frequency G-band spectral power of our State 2 over that of State 1 (Figure 4) actually begins at 10 mHz and is fully present by 30 mHz. We can just reach this range in a few Hinode data sets, and we find there the same association of spectral power with MBPs as here. This result would not be affected by terrestrial seeing.

We encountered two distinct states of the solar atmosphere (Figure 4). State 1 defined by $I_G = 0.6 - 1.4$ contained 98.6% of the data. It showed relatively low mean $I_K$ and UHF spectral power 0.. State 2 defined by $I_G \geq 1.65$ contained only 0.15% of the data showed strongly increased mean $I_K$ and mean UHF spectral power. Previous work (Schrijver 1989; Rezaei et al. 2007; Loukitcheva 2009) found that network features showed a power law relation between the photospheric magnetic field strength and the intensity of Ca II K-line emission. In contrast, we find a definite transition between the two "states" that occurs in the range $1.4 \leq I_G \leq 1.65$. This same transition is found in Hinode/SOT-FG images of internetwork. The CaII H-line intensity undergoes a step increase as G-band intensity increases past a critical value. In the case of network, however, after the step increase, the H-line intensity resumes increasing as G-band intensity increases.

For data filtered to the 3 - 5 mHz frequency band the kurtosis κ indicates a harmonic signal. For the UHF frequency band κ corresponds to an intermittent time series as illustrated in Figure 4. This provides further evidence that the UHF signal is neither uncorrelated noise nor harmonic oscillations but rather the result of a chaotic or turbulent process. The amplitude of the UHF fluctuations in State 1 is small compared to State 2, but the signal is still present and has a greater kurtosis.

Superposed epoch analysis showed that for both States 1 and 2 the UHF power of the $I_G$ fluctuations peaked simultaneously with $I_G$ itself. For State 1 $I_K$ shows a 3.5 minute oscillation and has a maximum 21 sec after the $I_G$ maximum. In the case of State 2, the $I_K$ shows no oscillation and its peak appears simultaneous with that of $I_G$.

In State 1 the $I_G$ and $I_K$ fluctuations give information on both the turbulent driving motions and the propagating acoustic wave. A time delay of 21 s and a velocity of 7 - 10 km/s imply a height separation of 150 - 210 km between the G-band and K-line sources. Using data obtained with the Interferometric Bidimensional Spectrometer at DST and assuming the 1-D FALC model of a static solar atmosphere (Fontenla et al. 1991), Reardon, Uitenbroek & Cauzzi (2009) found that the Ca II K-line response function peaks near height 400 km. Rimmele (2004), using the Universal Birefringent Filter at the DST, found that the G-band forms in the mid-photosphere near 200 km. Thus, our K-line observations appear dominated by an upper photospheric signal. For State 2 no delay appears between the G-band and K-line maxima. In this active, highly magnetized environment the geometrical configuration of the atmosphere may be sufficiently disturbed that sites of G-band and K-line emission lie spatially close together. This is an open question, of course, and requires further investigation.

Taken together, the above results: scale invariance, nonlinearity, and intermittence of the UHF G-band fluctuations, evoke the model described by Hasan et al. (2005) and Hasan & van Ballegooijen (2008) in which magneto-acoustic waves with frequencies above 10 mHz are generated by the interaction of flux tubes with turbulent downflows in the intergranular lanes.

Observations were obtained at the National Solar Observatory, operated by the Association of Universities for Research in Astronomy, Inc (AURA) under agreement with the National Science Foundation. DJC thanks California State University, Northridge for startup funds.

## REFERENCES

Anderson, L. S. & Athay, R. G. 1989, ApJ, 346, 1010

Bak, P., Tang, C. & Wiesenfeld, K. 1987, Phys. Rev. Letters, 59, 381

Bello Gonzalez, N., et al. 2010, ApJ., 723, L134

Carlsson, et al. 2007, PASJ, 59, S663

DeForest, C. E. 2004, ApJ, 617, L89

Fontenla, J. M., Avrett, E. H., & Loeser, R. 1991, ApJ, 377, 712

Fossum, A. & Carlsson, M, 2006, ApJ, 646, 579

Hasan, S. S., van Ballegooijen, A. A., Kalkofen, W. & Steiner, O. 2005, ApJ, 631, 1270-1280

Hasan, S. S. & van Ballegooijen, A. A. 2008, ApJ, 680, 1542

Jess, D. B., Mathioudakis, M., Christian, D. J., et al. 2010, Solar Phys., 261, 363

Jess, D. B., Andic, A., Mathioudakis, M. Bloomfield, D. S. & Keenan, F. P. 2007, A&A, 473, 943

Lawrence, J. K., Cadavid, A .C., Ruzmaikin, A. A. & Berger, T. E. 2001, Phys. Rev. Letters, 86. 5894

Loukitcheva, M., Solanki, S. K. & White, S.M. 2009, A&A, 497, 273

Lühr, H., Rother, M., Iyemori, T., Hansen, T. L. & Lepping, R. P. 1998, Annales Geophysicae 16, 743

Press, W. H., Flannery, B. P, Teukolsky, S. A. & Vetterling, W. T. 1989, Numerical Recipes in C, Cambridge University Press

Reardon, K. P., Uitenbroek, H. & Cauzzi, G. 2009, A&A, 500, 1239

Rezaei, R., Schlichenmaier, R., Beck, C. A. R., Bruls, J. H. M. J., & Schmidt, W. 2007, A&A, 466, 1131

Rimmele, T. 2004, ApJ, 604, 906

Schrijver, C. J., Cote, J., Zwaan, C., & Saar, S. H. 1989, ApJ, 337, 964

Schüssler, M., Shelyag, S., Berdyugina, S., Vögler, A. & Solanki, S. K. 2003, ApJ, 597, L173

Torrence, C. & Compo, G. P. 1998, Bull. Amer. Meteor. Soc., 79, 61

Weigelt, G. & Wirnitzer, B. 1983, Opt. Lett., 8(7), 389

Wöger, F., von der Lühe, O., & Reardon, K. 2008, A&A, 488, 375

**FIGURES**

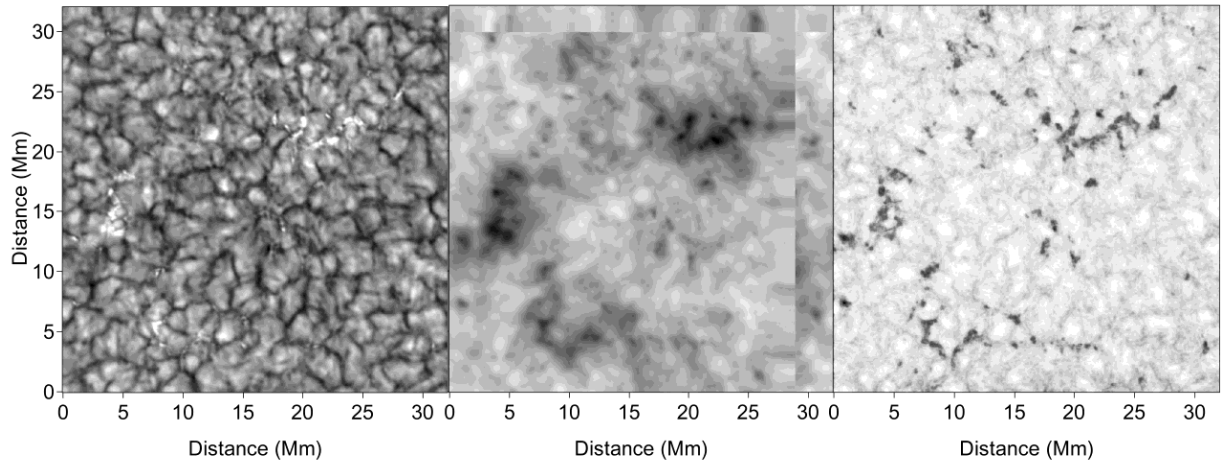

FIG. 1.—Time average of (left) the G-band intensity images, (middle) the CaII K-line intensity with the grayscale inverted, and (right) the spectral power of 25-100 mHz G-band fluctuations, also with grayscale reversed. The K-line image has been shifted to align with the G-band image.

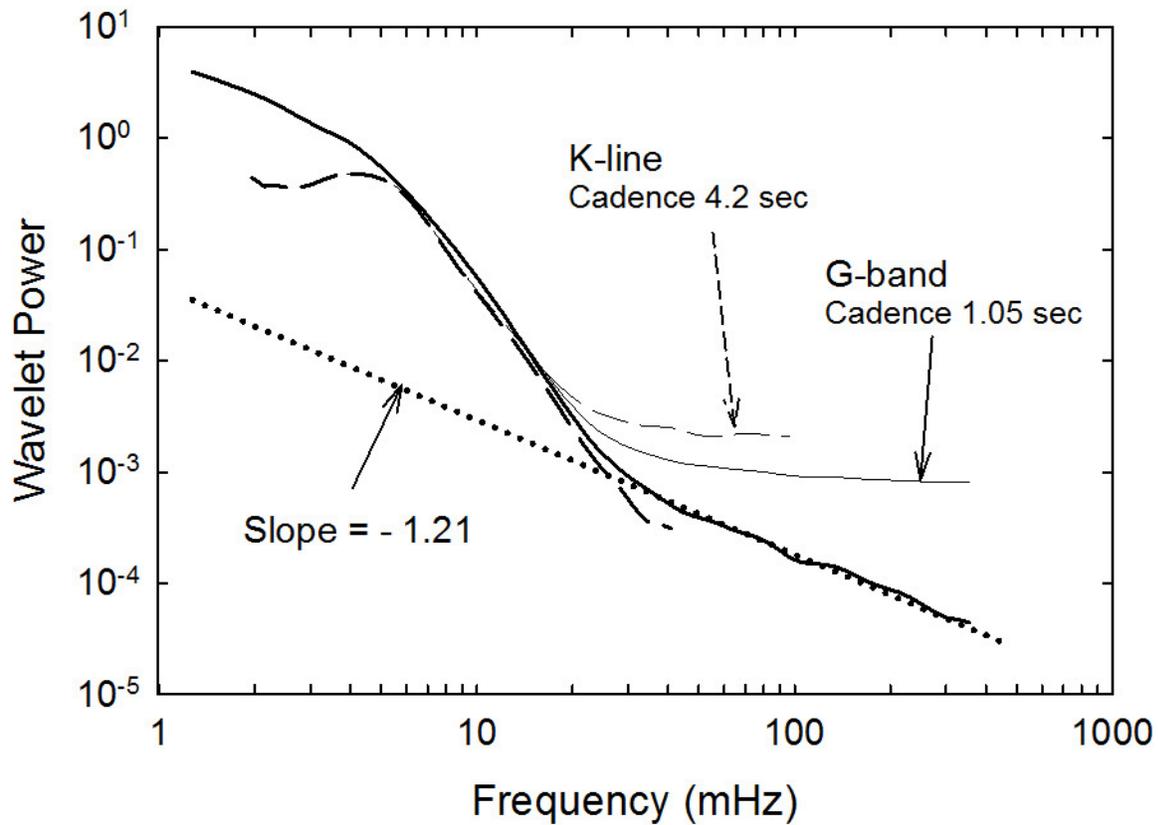

FIG. 2.— Morlet wavelet spectra averaged over space-time points throughout the data block. Thin solid line: raw G-band spectrum at cadence 1.05 sec. Thin dashed line: raw K-line spectrum at cadence 4.2 sec. The thick solid and dashed lines are the noise filtered spectra of G-band and K-line, respectively. The dotted line represents a best fit to the G-band power law for $f = 28 - 326$ mHz.

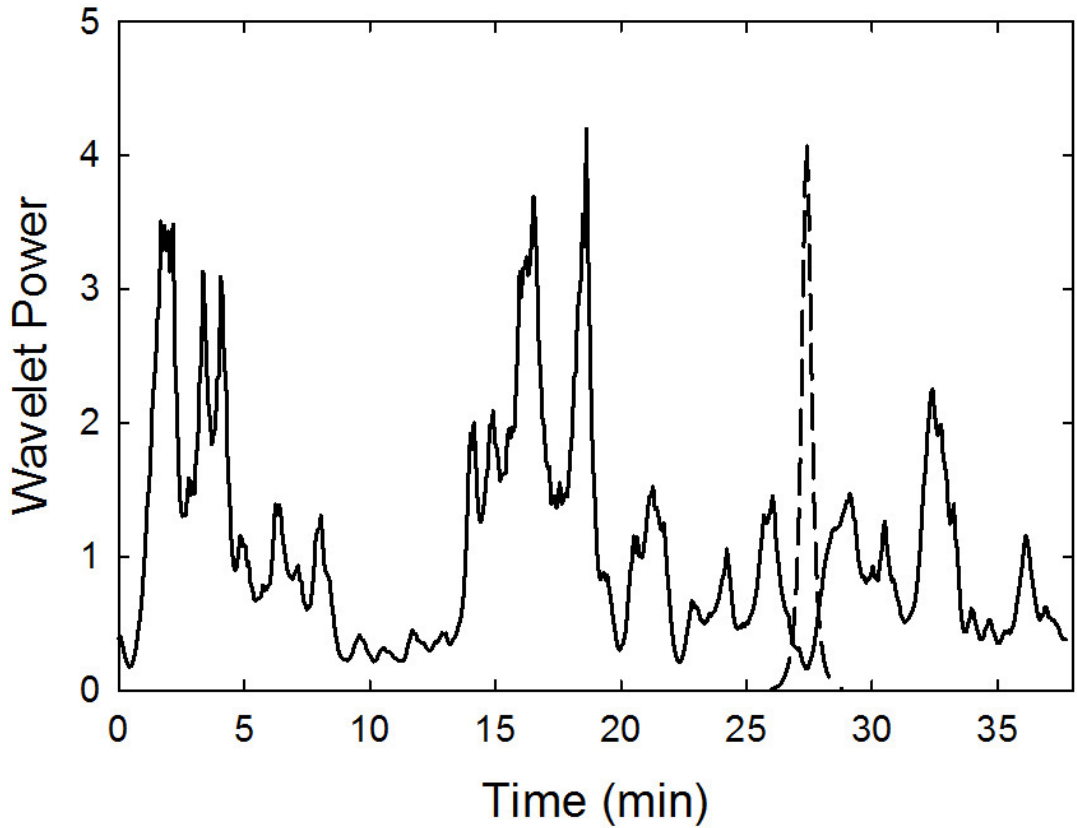

FIG. 3.— Time dependence of the UHF G-band spectral power at one spatial location corresponding to a G-band bright point and at 4.2 sec cadence (solid line). The dashed line is the spectral response to a delta-function intensity impulse.

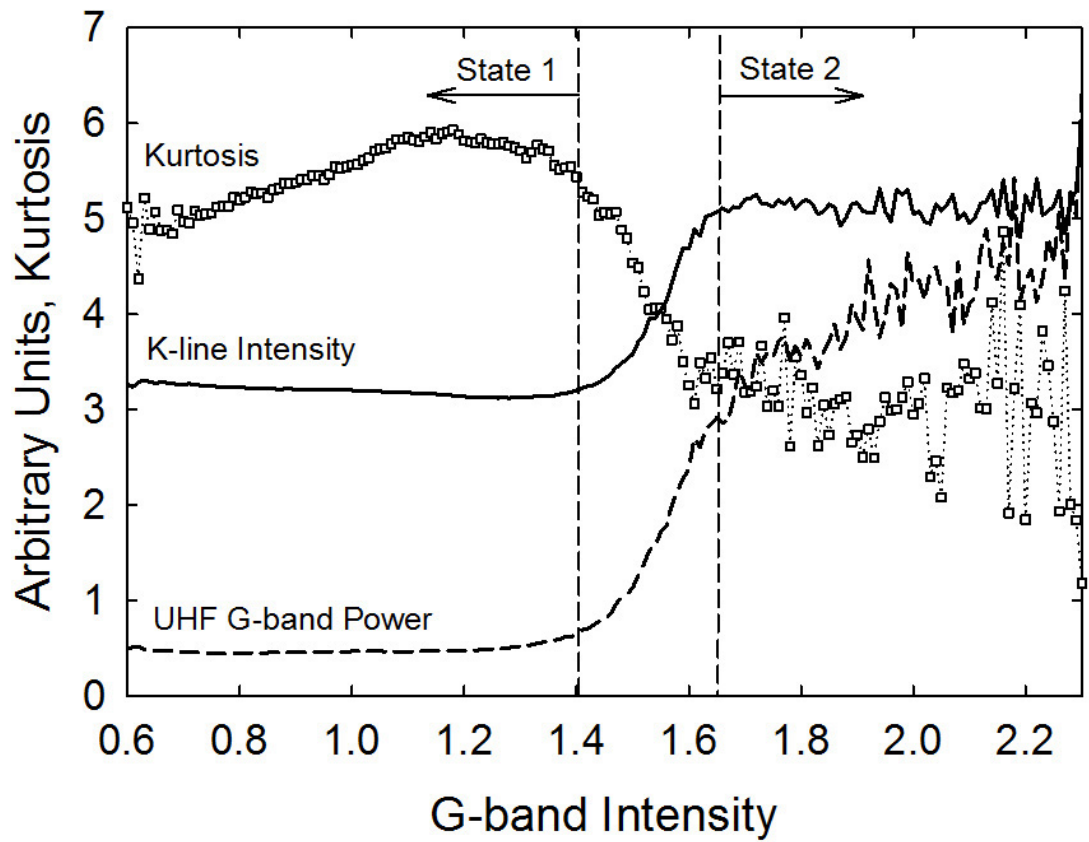

FIG. 4.— K-line intensity (solid), G-band UHF spectral power (dashed), and kurtosis of the G-band fluctuations filtered to the UHF frequencies (open squares) versus G-band intensity in bins of 0.01. The G-band and K-line intensities and the UHF power are in arbitrary units. The kurtosis is a dimensionless number. The sizes of local error bars are well indicated by the short-term scatter in the plots.

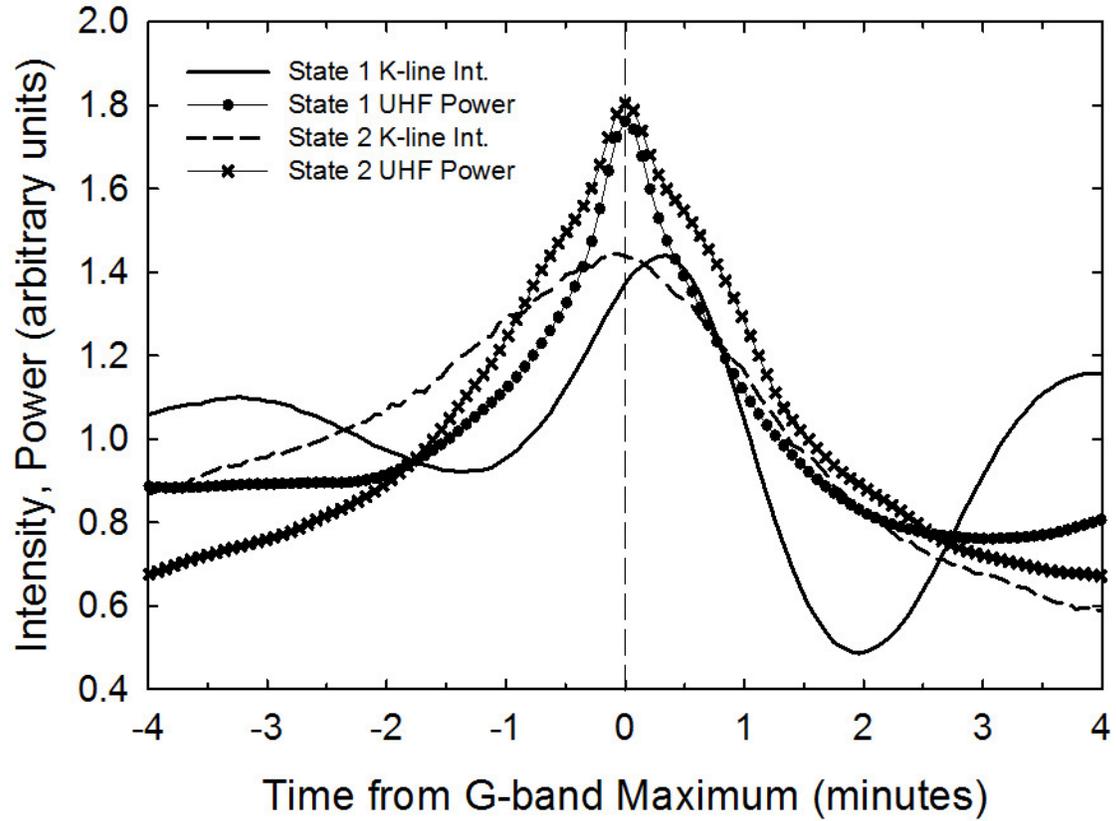

FIG. 5.— Averaged plots of $I_K$ and G-band UHF power zeroed on times of maximum $I_G$. The solid curve and the filled circles correspond to the requirement that the maxima belong to State 1 and represent $I_K$ and G-band UHF power, respectively. The dashed line and x's represent the same for State 2. The units of all the variables are arbitrary and have been adjusted for clarity of display.